\documentstyle[12pt]{article}
\makeatletter
\if@twoside
\else \oddsidemargin 00pt \evensidemargin 00pt
\fi
\let\@eqnsel = \hfil

\expandafter\ifx\csname mathrm\endcsname\relax\def\mathrm#1{{\rm #1}}\fi
\@ifundefined{mathrm}{\def\mathrm#1{{\rm #1}}}{\relax}

\makeatother

\begin{document}
\thispagestyle{empty}
\null
\hfill CERN-TH.7127/93
\vskip 1cm
\vfil
\begin{center}
{\Large \bf Can EROS/MACHO be detecting the galactic spheroid
instead of the galactic halo ?
\par} \vskip 2.5em
{\large
{\sc G.F.\ Giudice\footnote{\rm On leave of
absence from INFN, Sezione di Padova.}}, {\sc S.\ Mollerach},
and {\sc E.\ Roulet} \\[1ex]
{\it Theory Division, CERN, Geneva, Switzerland} \\[2ex]
\par} \vskip 1em
\end{center} \par
\vskip 4cm
\vfil
{\bf Abstract} \par

Models of our galaxy based on dynamical observations predict
a spheroid component much heavier than accounted for by direct
measurements of star counts and high velocity stars. If, as first
suggested by Caldwell and Ostriker, this discrepancy is due to a
large population of faint low-mass stars or dark objects in the
spheroid, the spheroid could be responsible for microlensing events
for sources in the Large Magellanic Cloud (LMC).
We show that, although the rate of events is
lower than predicted by a galactic halo made of microlensing objects,
it is still significant for EROS/MACHO observations. Because of the
different matter distributions in the halo and spheroid components,
a comparison between microlensing event rates in the LMC, future
measurements of microlensing in the galactic bulge and, possibly, in
M31 can provide information about the amounts of
dark objects in the different galactic components. If the EROS/MACHO
collaborations find a deficiency with respect to their halo
expectation, when more statistics are available, their detected events
could be interpreted as coming from spheroid microlenses, allowing for
a galactic halo composed entirely of non-baryonic dark matter.
\par
\vskip 2cm
\noindent CERN-TH.7127/93 \par
\vskip .15mm
\noindent December 1993 \par
\null
\setcounter{page}{0}
\clearpage

\section{Introduction}

The recent detection by the EROS \cite{eros} and MACHO \cite{macho}
collaborations of microlensing events in the Large Magellanic Cloud
(LMC) seems to support the hypothesis that at least part of the dark
matter in our galaxy is baryonic, appearing in the form of massive
astrophysical objects. Indirect evidence for the presence of galactic
dark matter existed from several observations like measurements of the
rotation curves of spiral galaxies, including our own, or velocity
dispersions measured in elliptical galaxies.

At present, all dynamical observations in our galaxy are well fitted
by models with at least three galactic components. These are: {\it (i)}
the (relatively thin and flat) {\it disk}, with a surface
mass density decreasing
exponentially with $r$, the distance from the galactic centre,
{\it (ii)} the (approximately spherical)
{\it halo}, with
density asymptotically falling as $r^{-2}$, which gives the main
contribution to the total mass of the Galaxy and ensures the flatness
of the rotation curve at large distances, and {\it (iii)} the
(approximately spherical) {\it
spheroid}, which contributes sizeably to the central part of the galaxy
but has a faster radial decrease than the halo. In some models,
additional components are also introduced.

The total mass in spheroid stars with mass $m$ larger than the minimum
mass for hydrogen burning was estimated, on the basis of star counts at
high galactic latitudes and of high velocity stars by Bahcall,
Schmidt
and Soneira \cite{bss}, to be $M_S(m>0.085M_\odot ) =0.9-3.2 \times
10^9 M_\odot$. On the other hand, the galactic models based on
dynamical measurements and on 2.2~$\mu$m infrared (IR) surveys of the
galactic centre predict a much larger total mass for the spheroid,
$M_S =5$--$7\times 10^{10}M_\odot$, which is comparable to the mass of
the disk \cite{co,oc,rk}. As recognized by Caldwell and Ostriker
\cite{co} and by Bahcall, Schmidt and Soneira \cite{bss}, these two
results can be made compatible by assuming that most of the mass of
the spheroid is non-luminous, in the form of faint low-mass stars,
neutron stars, brown dwarfs or Jupiters. An alternative solution
\cite{bss} to this apparent discrepancy is to consider a light
spheroid, as determined by star counts and high velocity stars,
together with a new galactic component, a central {\it core} with
large mass density and a sharp cut-off at about 1 kpc, which accounts
for the innermost galactic observations. One should note however that
there is no reason for the stellar mass function to become negligible
just below the hydrogen burning limit. In fact, recent measurements of
spheroid field stars \cite{fahlman} indicate a steeply increasing mass
function towards low masses, with no indications of flattening or of a
cutoff near 0.1 $M_\odot$, and this provides support for the heavy
spheroid models constituted mainly by brown dwarfs.

We consider here the implications of the scenario with a heavy,
mostly dark, spheroid for the ongoing microlensing searches,
showing that it can give rise to a significant rate of events
both for stars in the LMC and in the galactic bulge. The paper
is organized as follows. In Sec. 2 we compare different galactic
models and give the corresponding density profiles for the spheroid
and the halo which will be used in our study. In Sec. 3 we discuss the
prediction for the microlensing event rate at the LMC and compare the
results with the EROS/MACHO data. Sections 4 and 5 are devoted to
microlensing in the galactic bulge and in M31, where further
signatures for spheroid dark objects can be found. Our conclusions are
drawn in Sec. 6.

\section{Galactic models and density profiles}

We will use four different galactic models which provide good fits to
the observed rotation curves and the results from 2.2 $\mu$m IR
surveys, and which predict a heavy spheroid. As stressed by Caldwell
and Ostriker the dynamical galactic measurements lead to a spheroid
component much heavier than accounted for by the brighter visible stars
\cite{co}.
We use their best-fit models C(150) and D(150) \cite{oc}, which we
will call OC1 and OC2. Model OC1 includes in the fit the inner peak
in the rotation curve and not the measured bulge velocity dispersion,
while model OC2 does the reverse. These models consist of three
components. The disk has an exponentially falling surface density, and
the spheroid has a density (obtained from a deconvolution of the
``Hubble Law")
\begin{equation}
\rho_S^{\mbox{\scriptsize (OC)}}(r) = \rho_S \left\{ \begin{array}{ll}
{3.75\over Z^2}\left({3-Z \over Z^{1/2}} \ln { 1+Z^{1/2}\over (1
-Z)^{1/2}}-3 \right), & \mbox{$r < r_s$}\\
{3.75\over Z^2}\left({3+Z \over Z^{1/2}} \left[\arctan\left(
 {- 1 \over Z^{1/2}}\right)+{\pi \over 2} \right] -3 \right), &
\mbox{$r > r_s$}
\end{array} \right.
\end{equation}
where $Z=|(r/r_S )^2-1|$. In model OC1, $r_S=0.09929$ kpc, $\rho_S=
136.0 M_\odot$ pc$^{-3}$, while, for the model OC2, $r_S=0.1004$ kpc,
$\rho_S=101.8 M_\odot$ pc$^{-3}$.
The asymptotic behaviour at large
radius is $\rho_S^{\mbox{\scriptsize (OC)}}\sim r^{-3}$ and the total spheroid
masses are $M_S^{(OC1)}=5.9
\times 10^{10}M_\odot$, and $M_S^{(OC2)}=4.6
\times 10^{10}M_\odot$, assuming a density cut-off at 150 kpc.
Finally, there is a dark halo with density
\begin{equation}
\rho_H^{\mbox{\scriptsize (OC)}}(r)=\frac{\rho_c}{1+(r/r_c)^2},
\label{roc}
\end{equation}
where $r_c=3.697$ kpc, $\rho_c=7.815\times 10^{-2}
M_\odot$ pc$^{-3}$ for OC1 and $r_c=2.332$ kpc,
$\rho_c=20.03\times 10^{-2}
M_\odot$ pc$^{-3}$ for OC2.

We will also consider the models 5 and 6 of Rohlfs and Kreitschmann
\cite{rk}, here called RK1 and RK2, which are also based on dynamical
observations and 2.2 $\mu$m IR surveys, and predict a heavy spheroid
as well. The two models differ only in that in model RK1 all data
points of the rotation curves are included, whereas in model RK2, the
data for 1  ${\rm kpc}<r<3$ kpc are omitted. Here the spheroid follows a
spherical Brandt profile
\begin{equation}
\rho_S^{\mbox{\scriptsize (RK)}}(r)=\rho_b\left[
1+2(r/r_b)^{n_b}\right]^{-(1+3/n_b)},
\end{equation}
where $r_b=0.68$ kpc, $n_b=0.54$, $\rho_b=2.59\times 10^{3}
M_\odot$ pc$^{-3}$ for RK1 and $\rho_b=2.25\times 10^{3}
M_\odot$ pc$^{-3}$ for RK2, with a total mass $M_S^{(RK1)}=7.2
\times 10^{10}M_\odot$ and $M_S^{(RK2)}=6.3
\times 10^{10}M_\odot$. In their models the halo has the density
\begin{equation}
\rho_H^{\mbox{\scriptsize (RK)}}(r)=\rho_h\frac{1+n_hx^{-n_h}}{x^2}\exp
(-x^{-n_h}),
\label{rkh}
\end{equation}
where $x=r/r_h$, $r_h=14.53$ kpc, $n_h=3.40$, $\rho_h=3.27\times
10^{-3} M_\odot$ pc$^{-3}$ for RK1 and $r_h=14.64$ kpc, $n_h=3.00$,
$\rho_h=3.55\times 10^{-3} M_\odot$ pc$^{-3}$ for RK2. These models
contain also a central core with constant density $\rho_k$ and cut-off
$r_k$, where $\rho_k=27 M_\odot$ pc$^{-3}$ and $r_k=0.237$ kpc for RK1
and $\rho_k=11 M_\odot$ pc$^{-3}$ and $r_k=0.390$ kpc for RK2.

The general structure of RK1 and RK2 is substantially similar to OC1
and OC2, for which the central core is mimicked by a logarithmic rise
in $\rho_S^{\mbox{\scriptsize (OC)}}(r)$ at
small $r$. The main difference is the halo
density minimum of RK1 and RK2 at the galactic centre, which however
plays no significant role in our discussion.

In Ref. \cite{rk}, models with lighter spheroids are also proposed.
However, for comparison, we will consider the galactic model of
Bahcall, Schmidt, and Soneira \cite{bss} (here called BSS) for which
the density profile of the spheroid is determined by non-dynamical
observations. BSS is a four-component model with a spheroid having a
``de Vaucouleurs" density
\begin{equation}
\rho_S^{\mbox{\scriptsize (BSS)}}(r)= {\rho_D(R_0)\over 800} \left\{
\begin{array}{ll}
1.25 Y^{-3} \exp\left[10.093(1 - Y) \right], & \mbox{$r < 0.03 R_0$}\\
Y^{-7/2}\left(1-{0.08669 \over Y}\right) \exp\left[10.093(1-Y)\right] ,
 & \mbox{$r \geq 0.03 R_0$},
\end{array} \right.
\end{equation}
with $Y \equiv (r/R_0)^{1/4}$, $\rho_D(R_0) = 0.15 M_{\odot}$
pc$^{-3}$ and $R_0 = 8$ kpc, the galactocentric solar distance, and
with total mass $M_S^{\mbox{\scriptsize (BSS)}}=4.2\times 10^9 M_\odot$.
The halo density in their model is
given by
\begin{equation}
\rho_H^{\mbox{\scriptsize (BSS)}}(r)=\rho_H(R_0) \left({a^{1.2} +R_0^{1.2}
\over
a^{1.2} +r^{1.2}}\right)\times \left\{ \begin{array}{ll}
1, & \mbox{$r \leq R_C$}\\
(R_C/r)^{1.5}, & \mbox{$r > R_C$\ ,}
\end{array}
\right.
\end{equation}
with $\rho_H(R_0) = 0.009 M_{\odot}$ pc$^{-3}$, $R_C =30$ kpc, and
$a =2$ kpc. Besides the disk, there is also a central core, which is
important at $r~\raisebox{-.4ex}{\rlap{$\sim$}} \raisebox{.4ex}{$<$}~
1$~kpc.

The density profiles for halo and spheroid
in the different models are compared in Fig.~1. Notice the clear
discrepancy between the spheroid mass density in models OC and RK
and in the BSS model.

We will assume, as suggested by Caldwell and Ostriker \cite{co} and by
Bahcall, Schmidt and Soneira \cite{bss}, that the difference between
the predictions of the spheroid mass based on dynamical observations
and that based on star counts and
high-velocity stars is due to a large spheroid
population of very faint stars or dark objects. Given the large
numerical discrepancy between the two values of $M_S$, we will take
$\rho_S(r)$ in the heavy spheroid models as describing the density of
microlensing objects to a good approximation and proceed to estimate
the predicted event rates for EROS/MACHO.

\section{Microlensing in the LMC}

The optical depth $\tau$ for microlensing, which is just the number of
objects inside the microlensing tube, can easily be computed
following Paczy\'nski \cite{pac}:
\begin{equation}
\tau =\frac{4 \pi G u_T^2}{c^2 L}
\int_0^L dx x (L - x) \rho(r).
\label{tau}
\end{equation}
In Eq.~(\ref{tau}) $L$ is the distance to the source, $x$ is
the distance to the lensing object, $u_T$ is
the experimental threshold for $u\equiv d/R_e$, where $d$ is the
impact parameter of the lensing object, i.e. its
minimum distance from the line-of-sight and
\begin{equation}
R_e=2\sqrt{\frac{Gm}{c^2}\frac{(L-x)x}{L}}
\end{equation}
is the Einstein radius for an object with mass $m$. Since $u$ is
related to the image amplification $A$ by the relation \cite{vie}
\begin{equation}
A=\frac{u^2+2}{u\sqrt{u^2+4}},
\end{equation}
by choosing, e.g., $A_T=1.34$ one finds $u_T=1$.
In Eq.~(\ref{tau}), the distance of the lensing object to the
galactic centre is
\begin{equation}
r=\sqrt{x^2+R_0^2-2xR_0\cos b \cos l}
\end{equation}
with $(b,l)$ the angular galactic coordinates of the source (we have
taken $R_0=8.5$ kpc except for the BSS model).

Figure 2 shows the optical depth $\tau$ as a function of $\cos b
\cos l$, for three different source distances $L$ (corresponding
to the distance to
the galactic centre, the LMC, and M31)  and for microlensing
objects distributed as the spheroid and halo density profiles
of the OC1 model. Notice that, as $\cos b \cos l$ increases and
the line-of-sight comes closer to the galactic centre, the
optical depth for spheroid lensing objects becomes larger and
can overcome the value of $\tau$ from halo lensing objects.
Also, as $L$ increases, the optical depth for spheroid objects
saturates much faster than for halo objects, because of the
steeper decrease with the radius of the spheroid density.

For the LMC ($L=55$ kpc, $b=-32.8^\circ$, $l=281^\circ$), the optical
depths for the spheroid models considered here, as well as the
prediction for a halo of lensing objects, are given in Table~1.

To compute the rate of events $\Gamma$, which corresponds to the flux
of objects entering the microlensing tube, it is necessary to know the
velocity dispersion of the spheroid population. This has been measured
by Hartwick and Sargent \cite{har} at distances of about 10 kpc and 60
kpc and found a value of $\sigma$, the one-dimensional velocity
dispersion, $\sigma\simeq$ 125~km/s. Note that for an isothermal and
isotropic velocity distribution with a mass density $\rho \sim
r^{-n}$, the velocity dispersion is related to the circular velocity
$v_c$ through $\sigma =v_c/\sqrt{n}$ \cite{har,rich90} giving, for
$\sigma =$ 125~km/s and $v_c =220$~km/s, $n=3.1$, in agreement with
the asymptotic behaviour of the models considered ($n=3$ for OC,
$n=3.54$ for RK).

Following Griest \cite{gri}, we compute $\Gamma$ from the expression
\begin{equation}
\Gamma =8u_T \sigma \sqrt{\frac{2G}{c^2mL}}\int_0^L dx
\sqrt{x(L-x)}\rho (r) e^{-\eta^2}\int_0^\infty dz e^{-z^2}z^2
I_0(2z\eta ),\label{gam}
\end{equation}
where $I_0$ is the modified Bessel function of order 0. In Eq.
(\ref{gam}) $\eta =v_t(x)/(\sqrt{2}\sigma )$ and $v_t(x)$ is the
transverse velocity of the microlensing tube at a distance $x$ given by
\begin{equation}
v_t(x)=\sqrt{\left(1-{x\over L}\right)^2|{\vec{v}}_{\odot_\perp}|^2+
\left({x\over L}\right)^2|{\vec{v}}_{s_\perp}|^2
+2{x\over L}\left(1-{x\over L}\right)
|{\vec{v}}_{\odot{\perp}}||{\vec{v}}_{s_\perp}| \cos \theta },
\end{equation}
where ${\vec{v}}_{s_\perp}$ and ${\vec{v}}_{\odot_\perp}$ are the source
and solar velocities transverse to the line-of-sight, and $\theta$ is
the angle between them.

The values of $\Gamma$ for the LMC are shown
in Table~1 for the different models. To allow easier comparison with
the existing literature, we also present the predicted rates in the
case of a halo made of  lensing objects, in which we have chosen
$\sigma =155$~km/s (corresponding to a 3-dimensional velocity
dispersion of 270~km/s).
Notice that the predicted halo event rates vary by as much as a factor
of 1.7 among different models. In particular, the rate quoted in
Ref.~\cite{gri} lies in the lower end of the range shown in Table~1.

The differential rate distribution in event duration $t$ is
\begin{equation}
\frac{d\Gamma}{dt}=\frac{8 u_T \sigma^2}{m}
\int_0^L dx
\rho (r) z^4 e^{-(z^2+\eta^2)} I_0(2\eta z) ,
\end{equation}
with
\begin{equation}
z\equiv \sqrt{\frac{2Gmx(L-x)}{c^2L\sigma^2 t^2}}.
\end{equation}
The event time duration $t$ is defined as the time taken by the dark
object to travel a distance equal to $R_e$ in a direction orthogonal
to the line-of-sight. We can compute the average time duration
\begin{equation}
\langle t\rangle \equiv \int_0^\infty
dt~\frac{t}{\Gamma}\frac{d\Gamma}{dt}=\frac{2\tau}{\pi
u_T\Gamma},
\end{equation}
which is given in Table~1.
One can also estimate the most probable event
duration as the time $t_P$ at which $\Gamma^{-1}d\Gamma /dt$ reaches a
maximum.
The values of $t_P$ for the different models are shown in Table~1.

For a given time duration $t$, we can compute the most probable mass
$m_P$ of the object responsible for that microlensing event as the
value which maximizes the distribution $\Gamma^{-1}d\Gamma /dt$,
taken now as a function of $m$.
The results for $m_P$, shown in Table~1, can only be understood as
indicative,
and masses three times smaller or three times larger are roughly half
as probable.
For large statistics, information on the lensing
object mass distribution could be extracted using the technique
proposed in Ref. \cite{der}.
Notice that the results contained in Table~1 have
been derived under the simplifying assumption that all lensing objects
have the same mass.

The expected number of events is then simply
\begin{equation}
N_{ev}=\Gamma N_\star T \epsilon
\end{equation}
with $N_\star$ the number of monitored stars, $T$ the total observation
time, and $\epsilon$ the experimental efficiency. For the microlensing
events due to dark
objects of the spheroid, we estimate from Table~1 :
\begin{equation}
N_{ev}=0.16 u_T\epsilon \sqrt{\frac{M_\odot}{m}}{N_\star\over 10^6}
{T\over{\mbox{years}}},
\label{neve}
\end{equation}
where we have chosen the most favorable case of model RK1.
Results from the other models can be easily extracted from Table~1.

MACHO has observed one event with a duration of 17 days and EROS
has observed two events with durations
of 27 and 30 days. From the results of Table~1 we estimate the most
likely mass range $2 \times 10^{-2} < m/M_\odot < 2 \times 10^{-1}$
for $t=15$ days, and
an interval of masses four times heavier for $t = 30$ days.
For the MACHO collaboration, which has monitored 1.8 million stars
during one year, the expected number of events is about $N_{ev}= 1
\sqrt{0.1 M_\odot/m}~u_T \epsilon$, while for EROS, with 3 million
stars
monitored during three observing seasons of about six months each,
the expected number is about $N_{ev}=2.2
\sqrt{0.1 M_\odot/m}~u_T \epsilon$.

The experimental efficiency $\epsilon$ can only be computed with a
specific knowledge of the experimental apparatus and of the details of
the observational procedures, and has not been fully presented by the
two groups. The predictions from spheroid microlensing seem to be lower
than the preliminary observations of EROS/MACHO, unless either their
efficiencies are not much smaller than 1 or the lensing objects are
lighter than what suggested by the observed event durations.
Only with improved
experimental statistics and a careful study of the efficiencies can
more conclusive statements be made. Nevertheless, the rate in
Eq.~(\ref{neve}) is still significant and could allow for a
determination of the amount of spheroid dark matter, especially if
observations from different sources are compared, as discussed in the
next section.

Finally we want to mention that a larger event rate could be expected
if the LMC contained a spheroid of dark objects similarly to the Milky
Way. Since the LMC is an irregular galaxy, we cannot draw analogies
with our galaxy and we do not attempt to make any estimates.
We can only expect that, if microlensing events from LMC spheroid dark
objects indeed occur, they should be rather sensitive to the distance
of the line-of-sight with the centre of the LMC galaxy.

\section{Microlensing in the galactic bulge}

Since the galactic spheroid is highly concentrated towards the
galactic centre, where it becomes the dominant component, microlensing
of bulge (central spheroid) stars could be a strong test of the heavy
dark spheroid models that we are considering. This is particularly
interesting as the OGLE group \cite{ogle} is currently attempting to
detect microlensing of bulge stars and MACHO is planning to do so
during the southern winter, when the LMC is low. Microlensing of bulge
stars by the halo and disk dark matter and by faint stars in the disk
has already been studied by Paczy\'nski \cite{pacbul} and by Griest
et al. \cite{grbul}.

Due to the strong radial dependence of the spheroid density, both the
optical depth and the event rate are sensitive functions of the angle
between the source and the galactic centre. The optical depth has
already been given in Eq.~(\ref{tau}). The computation of the event
rate is analogous to the one that has led to Eq.~(\ref{gam}); here,
however, one has to average not only over the velocity distribution of
the lenses, but also over the velocity distribution of the
sources ${\vec{v}}_{s_\perp}$. The total event rate is
\begin{equation}
\Gamma_B =\frac{1}{2\pi \sigma_b^2}\int_0^{2 \pi}d\theta \int_0^\infty
dv_s v_s\exp\left(-\frac{v_s^2}{2\sigma_b^2}\right) \Gamma ,
\end{equation}
where $v_s \equiv |{\vec{v}}_{s_\perp}|$, $\Gamma$ is given in
Eq.~(\ref{gam}), and an isotropic Maxwellian velocity distribution has
been assumed. Since the velocity dispersion measured for bulge stars
is slightly smaller than for spheroid stars at larger radius \cite{har},
we take
for the one-dimensional velocity dispersion the value $\sigma_b =
105$~km/s, in accordance with the value obtained by Rich
\cite{rich90} for Baade's window ($l,b) = (0.9^\circ,
-3.9^\circ)$. We have neglected in our computations the effect of the
bulge rotation, which may affect the microlensing rates by spheroid
objects by as much as $\sim 20 \%$. However, the effect is more
important for lensing from disk dark objects \cite{grbul}.

The rates for bulge star microlensing from spheroid dark objects are
plotted in Fig.~3 as a function of the angle $\alpha$ between the
source and the galactic centre ($\alpha = 4^\circ$ for Baade's
window) for the spheroid models OC1 and RK1. We also show for
comparison the results for the halo dark matter in model OC1 and those
obtained in Ref. \cite{grbul} for disk dark matter (DDM). The rate
from halo dark matter is independent of the angle between the
line-of-sight and the galactic centre (for small angles) since, for
$r~\raisebox{-.4ex}{\rlap{$\sim$}} \raisebox{.4ex}{$<$}~r_c =3.7$~kpc,
$\rho_H^{\mbox{\scriptsize (OC)}}$ is roughly constant, see Eq.~(\ref{roc}).
For DDM, the angle coordinate in the plot is the absolute
value of the galactic latitude $b$ (and not $\alpha$) and the rate is
almost independent of the galactic longitude $l$ (for not too large
values of $l$). The existence of DDM claimed for many years \cite{ddm}
is still very controversial. A more established source of microlensing
comes from faint low-mass stars in the disk. The predictions for the
optical depth and the event rate are rather uncertain, depending on
the extrapolation of the mass distribution function \cite{grbul}.
Since the event rate is computed by integrating over all faint stars
masses, we do not show it in Fig.~3, where a fixed value of $m$ has
been assumed. Its angular dependence is qualitatively similar to the
one of DDM. The predictions for the optical depth, microlensing event
rate and average event duration for faint disk stars, DDM, halo and
spheroid dark matter are shown in Table~2 for stars in Baade's
window. The uncertainties in the fit of the faint disk stars mass
distribution, as computed in Ref.~\cite{grbul}, are shown in Table~2.
We have not considered the RK halos, since they predict a negligible rate,
because of their ``holes" in the galactic centre, see Eq.~(\ref{rkh})
and Fig.~1.

A realistic range accessible to observations is $\alpha \sim
4^\circ$--$10^\circ$, since the galactic centre is heavily obscured. As shown
in
Fig.~3, for $l \simeq 0$, the spheroid dark matter event rate is
comparable to, and even larger than, the event rates expected from
both DDM and halo dark matter. It should be emphasized that the halo
rate is very sensitive to the assumed value of the core radius,
decreasing for larger $r_c$.

As mentioned above, we have plotted in Fig.~3 the spheroid
microlensing event rate  as a function of $\alpha$, whereas the DDM
microlensing event rate is plotted as a function of $|b|$. The
different and very strong dependence on the galactic longitude $l$
provides the ``smoking gun" signature that allows one  to distinguish
spheroid microlensing events from those coming from DDM or faint disk
stars.

\section{Microlensing in the Andromeda galaxy}

Although M31 is much further away ($L=$ 650 kpc) than the Magellanic
Clouds, it has been suggested that microlensing of M31 stars could be
a good probe of the presence of baryonic dark matter around that
galaxy \cite{cro}. In particular, the large inclination of its disk
($75^\circ$ with respect to the face on position) makes the optical
depth of stars in the far side larger than that in the near side
since the line-of-sight crosses  a larger fraction of the halo. This
effect will be even more pronounced if the lensing objects belonged to
a spheroid population rather than to a halo one, because of the
steeper $r$-dependence of the spheroid density profile.

Recent fits to dynamical observations in M31 show the presence of a
significant spheroid besides the disk. In Ref. \cite{bra}, a ``de
Vaucouleurs" spheroid law was assumed, as resulting from luminosity
measurements \cite{wat}, and fits to the rotation velocities resulted
in a total spheroid mass $M_S \sim 5 -8 \times 10^{10} M_{\odot}$. Due
to the great similarity between our galaxy and M31, we will just use
the OC1 spheroid model (which has a comparable total mass) to estimate
the optical depth for microlensing of M31 stars as a function of the
impact parameter $d$, i.e. the distance between the
line-of-sight and the M31 galactic centre. This is done by using Eq.
(\ref{tau}) where now the distance $r$ between the lensing object and
the M31 galactic centre can be written as:
\begin{equation}
r=\sqrt{(L-x-d\cos\Phi\tan i)^2+d^2},
\end{equation}
where $L$ and $x$ are defined in Sec. 3. Here $i\simeq 75^\circ$ is
the inclination of the M31 disk, while $\Phi$ is the position angle
relative to the far minor axis. The resulting optical depth for
spheroid lensing objects is shown in Fig.~4 as a function of the
impact parameter, for stars located along the minor axis,  with
positive values of $d$ representing the far side ($\Phi=0$) and
negative ones the near side ($\Phi=\pi$).

There is no convincing evidence for a halo in M31. Nevertheless, for
comparison, we also show in Fig.~4 the optical depth for lensing
objects forming an M31 halo described by the OC1 halo, as well as the
contribution to the optical depth from a Milky Way spheroid and halo
populations described by the same models. In the computation of the
optical depth for Milky Way lensing objects, a cut-off of 150 kpc has
been used for both spheroid and halo mass densities.

It is apparent from Fig.~4 that the M31 spheroid microlensing is the
most significant for small impact parameters and has a very different
profile from the one of the halo, providing a possible observational
test. We note, however, that the proximity between the sources and the
lensing objects could make the associated Einstein radius quite small;
its angular extent can therefore be smaller than the angle subtended
by some very large source stars. This could be a drawback for the
microlensing of red giant stars by very light objects \cite{cro}.

\section{Conclusions}

Galactic models obtained by fitting dynamical observations predict a
spheroid component considerably heavier than is accounted for by
estimates of the brighter
visible stars, suggesting that a ``missing mass" problem
exists in the spheroid. If such a spheroid dark matter really exists,
it will certainly be baryonic. In this paper we have assumed,
following Caldwell and Ostriker, that the spheroid population consists
mainly of low-mass faint stars or dark objects and we have computed
the rate for microlensing events for stars in the LMC. This rate is
lower than the corresponding one for microlensing events coming from a
galactic halo of dark objects, but it is still significant for
observations of EROS/MACHO. We also note that, if the LMC has a heavy
spheroid component, this could give an additional considerable
contribution to the total rate of events, since the line-of-sight can
pass very close to the LMC galactic centre. The example of M31,
presented in Sec. 5, is illustrative of such an effect. Unfortunately,
poor knowledge of the properties of the LMC does not allow us a
reliable estimate.

It is very interesting that the rates for spheroid and halo
microlensing vary considerably depending on the source. Searches
towards the galactic centre seem particularly promising for
distinguishing between them, since the rate for spheroid microlensing
is larger than for halo microlensing and has a characteristic
dependence on the galactic longitude. The much more challenging
search in M31 can also provide important information,
since the rate for spheroid microlensing can be very large and can
present a strong dependence on the impact parameter, if M31 has indeed
a spheroid component similar to the one of the Milky Way, as suggested
by observations.

Therefore microlensing searches can map the distributions of dark
heavy objects contained in the different galactic components,
especially if results from different sources are compared. Knowledge
of the amount of dark matter in the spheroid is of great interest for
understanding the structure of our galaxy and for building reliable
galactic models. Finally, we want to mention that the dark spheroid
contributions to the total mass of the Universe are not cosmologically
very significant, since the mass of the spheroid is comparable to the
mass of the galactic disk.

While Turner \cite{tur} has suggested that the preliminary EROS/MACHO
data already point towards some deficiency with respect to the halo
expectation, EROS \cite{eros} has claimed that their result is
consistent with their expectation, and MACHO \cite{macho} has made no
statement about it. Only improved statistics and a complete study of
the experimental efficiencies (which has not been presented by the
EROS/MACHO groups in their papers describing the discovery events
\cite{eros,macho}) can resolve the question. However, if a deficiency
is indeed found, this could be interpreted as microlensing entirely
due to dark objects in a heavy spheroid rather than in the halo. In
this scenario, therefore, the galactic halo can consist entirely of
non-baryonic dark matter and the EROS/MACHO observations could be
reconciled with the presently favoured model of structure formation in
a critical Universe. This of course would also have important
consequences for the currently running experiments searching for halo
dark matter from nuclear recoil and from annihilation products. The
different distributions of the two dark matter components, with the
baryonic one more concentrated towards the galactic centre, could
naturally be accounted for by the dissipative processes undergone by
the baryons during the galaxy collapse. This is in contrast to the
picture proposed by Turner and Gates \cite{tur2}, in which the same
galactic component, the halo, is formed by a mixture of baryonic and
non-baryonic dark matter.

\bigskip

Note added: After submitting this paper we received a preprint by
Gould, Miralda-Escud\'e and Bahcall \cite{gmb} where they discuss
the possibility
that the microlensing events are caused by dark objects in a thick (or a
thin) disk.
\bigskip

We are greatly indebted to Alvaro de R\'ujula for discussions,
criticism, and encouragement. We thank K. Griest and J. Rich for
useful correspondence. One of us (G.F.G.) wishes to thank Giacomo
G. for charming and edifying conversations.
We dedicate this work to him and to Javier R..
The work of S.M.
was supported by a grant of the European Communities (Human Capital
and Mobility Programme). E.R. was partially supported by Worldlab.

\pagebreak

\begin{table}[p]
\begin{center}
\caption{The optical depth ($\tau$), the event rate ($\Gamma$),
the average event duration ($\langle t\rangle$), the most probable event
duration ($t_P$), and the most probable mass ($m_P$),
for microlensing events in the LMC. We have taken $\sigma = $
125~km/s for the spheroid and $\sigma =$ 155~km/s for the halo. The
lensing object mass $m$ is in units of $M_\odot$ and $t_{10}$ is the
event duration in units of 10~days. We have taken $u_T = 1$
and $\tau\propto u_T^2, \Gamma\propto u_T$.}

\begin{tabular}{|c||c|c|c|c|c|}
\hline
 & $\tau$ &$\Gamma$ & $\langle t\rangle$ & $t_P$ & $m_P$ \\
 & &[$\frac{\mbox{\scriptsize{events}}}{{\mbox{\scriptsize{yr}}} ~10^6
{\mbox{\scriptsize{stars}}}}$]&[days]&[days]&[$M_\odot$]\\
\hline
\hline
Spheroid OC1 & $4.0\times 10^{-8} $ & .13 $/\sqrt{m}$&
71 $\sqrt{m}$& 44 $\sqrt{m}$& .03 $t_{10}^2$\\
Spheroid OC2 & $3.1\times 10^{-8} $ & .10 $/\sqrt{m}$&
71 $\sqrt{m}$& 44 $\sqrt{m}$& .03  $t_{10}^2$\\
Spheroid RK1 & $4.9\times 10^{-8} $ & .16 $/\sqrt{m}$&
69 $\sqrt{m}$& 42 $\sqrt{m}$& .03  $t_{10}^2$\\
Spheroid RK2 & $4.2\times 10^{-8} $ & .14 $/\sqrt{m}$&
70 $\sqrt{m}$& 42 $\sqrt{m}$& .03  $t_{10}^2$\\
Spheroid BSS & $3.5\times 10^{-9} $ & .01 $/\sqrt{m}$&
66 $\sqrt{m}$& 40 $\sqrt{m}$& .04 $t_{10}^2$\\
Halo OC1     & $7.5\times 10^{-7} $ & 2.7 $/\sqrt{m}$&
65 $\sqrt{m}$& 43 $\sqrt{m}$& .04  $t_{10}^2$\\
Halo OC2     & $8.0\times 10^{-7} $ & 2.8 $/\sqrt{m}$&
66 $\sqrt{m}$& 42 $\sqrt{m}$& .04 $t_{10}^2$\\
Halo RK1     & $5.4\times 10^{-7} $ & 1.6 $/\sqrt{m}$&
77 $\sqrt{m}$& 49 $\sqrt{m}$& .03  $t_{10}^2$\\
Halo RK2     & $5.8\times 10^{-7} $ & 1.8 $/\sqrt{m}$&
76 $\sqrt{m}$& 49 $\sqrt{m}$& .03  $t_{10}^2$\\
Halo BSS     & $8.4\times 10^{-7} $ & 2.7 $/\sqrt{m}$&
71 $\sqrt{m}$& 46 $\sqrt{m}$& .03  $t_{10}^2$\\
\hline
\end{tabular}
\end{center}
\end{table}
%
%
\begin{table}[p]
\begin{center}
\caption{The optical depth ($\tau$), the event rate ($\Gamma$), and
the average event duration ($\langle t\rangle$), for
microlensing events in Baade's window. The results for disk
DM and faint disk stars are taken from Ref. [13]. The lensing
object mass $m$ is in units of $M_\odot$.
We have taken $u_T = 1$
and $\tau\propto u_T^2, \Gamma\propto u_T$. }
\vspace{2em}
\begin{tabular}{|c||c|c|c|}
\hline
 & $\tau$ &$\Gamma$ & $\langle t\rangle$ \\
 & &[$\frac{\mbox{\scriptsize events}}{{\mbox{\scriptsize yr}}~
 10^6 {\mbox{\scriptsize stars}}}$]&[days] \\
\hline
\hline
Spheroid OC1      & $5.0\times 10^{-7} $ & 6.1 $/\sqrt{m}$&
19 $\sqrt{m}$\\
Spheroid OC2      & $3.9\times 10^{-7} $ & 4.6 $/\sqrt{m}$&
19 $\sqrt{m}$\\
Spheroid RK1      & $6.2\times 10^{-7} $ & 7.1 $/\sqrt{m}$&
20 $\sqrt{m}$\\
Spheroid RK2      & $5.4\times 10^{-7} $ & 6.3 $/\sqrt{m}$&
20 $\sqrt{m}$\\
Spheroid BSS      & $5.0\times 10^{-8} $ & 0.6 $/\sqrt{m}$&
21 $\sqrt{m}$\\
Halo OC1          & $2.7\times 10^{-7} $ & 2.4 $/\sqrt{m}$&
26 $\sqrt{m}$\\
Halo OC2          & $4.4\times 10^{-7} $ & 4.0 $/\sqrt{m}$&
25 $\sqrt{m}$\\
Halo BSS          & $1.3\times 10^{-7} $ & 1.2 $/\sqrt{m}$&
26 $\sqrt{m}$\\
Disk DM           & $9.8\times 10^{-7} $ & 5.1 $/\sqrt{m}$&
46 $\sqrt{m}$\\
Faint Disk Stars  & $(2.9-9.6)\times 10^{-7} $ &$ 2.2 - 7.5$ &
30 \\
\hline
\end{tabular}
\end{center}
\end{table}

\pagebreak

\noindent {\large{\bf Figure captions}}
\bigskip\bigskip

\noindent
{\bf Fig. 1}. Spheroid and halo density distributions
as a function of the radius $r$, for the various models
considered. Model RK2 (not shown) is very similar to RK1.

\bigskip

\noindent
{\bf Fig. 2}. Optical depth as a function of $\cos b \cos l$, with
$(l,b)$ the angular galactic coordinates, for three
different values of $L$, the
distance to the source, corresponding to the galactic centre
($L=8.5$ kpc), to the LMC ($L=55$ kpc) and to M31
($L=650$ kpc). Solid lines refer to spheroid lensing objects while
dashed lines refer to halo objects, as described by model OC1.
The locations of LMC and M31 are indicated.

\bigskip

\noindent
{\bf Fig. 3}. Microlensing event rates ($\times \sqrt{m/M_\odot}$)
expected for
bulge star sources. For the spheroid and the halo, the rate is plotted as a
function of $\alpha$, the angle between the source and the
galactic centre, which is the only relevant angle because of the
spherical symmetry of the density distribution. For disk dark matter,
the rate is plotted as a function of
the absolute value of the
galactic latitude $b$ and the rate is
independent of the galactic longitude $l$, in the range of interest.
The location
of Baade's window (BW) is indicated.

\bigskip

\noindent
{\bf Fig. 4}. Optical depth for microlensing of stars along the
minor axis of M31, as a function of the impact parameter $d$
($d>0$ for the far side and $d<0$ for the
near side). Solid lines refer to spheroid lensing objects and
dashed lines to halo objects, with the
contribution from M31 and the Milky Way (MW) plotted
separately.
The OC1 model was used.

\end{document}